\documentclass[journal]{IEEEtran}
\usepackage{stackengine}
\usepackage{blindtext}
\usepackage{graphicx}
\usepackage[utf8]{inputenc}
\usepackage{graphics} 
\usepackage{epsfig} 
\usepackage{mathptmx} 
\usepackage{amsmath} 
\usepackage{amssymb}  
\usepackage{cite}
\usepackage{multicol}
\usepackage{mathtools, cuted}
\usepackage[cmintegrals]{newtxmath}
\usepackage{epstopdf}
\usepackage{graphics} 
\usepackage{epsfig} 

\usepackage{mathptmx} 
\usepackage{amsmath} 
\usepackage{amssymb}  
\usepackage{cite}
\usepackage{multicol}
\usepackage{mathtools, cuted}
\usepackage[cmintegrals]{newtxmath}
\usepackage{mathtools}

\usepackage{mathptmx}
\usepackage{helvet}
\usepackage{courier}
\usepackage{subfigure}
\usepackage{type1cm}
\usepackage{titlesec}
\usepackage{epsfig} 
\usepackage{mathptmx} 
\usepackage{amsmath} 
\usepackage{amssymb}  
\usepackage{cite}
\usepackage{multicol}
\usepackage{mathtools, cuted}
\usepackage[cmintegrals]{newtxmath}
\usepackage{makeidx}         
\usepackage{algorithm}
\usepackage{algpseudocode}

\usepackage{multicol}        
\usepackage[bottom]{footmisc}
\usepackage{caption}
\usepackage{nomencl}
\usepackage[usenames, dvipsnames]{color}

\usepackage{multicol}        
\usepackage[bottom]{footmisc}
\usepackage{caption}
\usepackage{titlesec}
\usepackage{nomencl}
\usepackage[usenames, dvipsnames]{color}
\usepackage{multirow}

\usepackage{amsthm}
 
\theoremstyle{definition}
\newtheorem{definition}{Definition}

\theoremstyle{theorem}

\theoremstyle{remark}

\theoremstyle{proposition}

\theoremstyle{corollary}

\theoremstyle{proof}

\theoremstyle{assumption}

\theoremstyle{property}

\theoremstyle{lemma}

\ifCLASSINFOpdf
\else
\fi
\begin{document}
%
\title{Safe Human-UAS Collaboration Abstraction}
%
%
%

\author{Hossein Rastgoftar
\thanks{{\color{black}H. Rastgoftar is with the Department
of Aerospace and Mechanical Engineering, University of Arizona, Tucson,
AZ, 85721 USA e-mail: hrastgoftar@arizona.edu.}}
}
%
%

\markboth{
}%
{Shell \MakeLowercase{\textit{et al.}}: Bare Demo of IEEEtran.cls for IEEE Journals}
%



\maketitle
\begin{abstract}
This paper studies the problem of safe human-uncrewed aerial system (UAS) collaboration in a shared work environment. By considering human and UAS as co-workers, we use Petri Nets to abstractly model evolution of shared tasks assigned to human and UAS co-workers. Particularly, the Petri Nets' ``places’’  represent work stations; therefore, the Petri Nets' transitions can  formally specify displacements between the work stations. The paper's first objective is to incorporate uncertainty regarding the intentions of human co-workers into motion planning for UAS, when UAS co-workers closely interact with human co-workers. To this end,  the proposed Petri Nets model uses ``conflict’’ constructs to represent situations at which UAS deals with incomplete knowledge about human co-worker intention. The paper's second objective is then to plan the motion of the UAS in a resilient and safe manner, in the presence of non-cooperative human co-workers. In order to achieve this objective, UAS equipped with onboard perception and decision-making capabilities are able to, through real-time processing of in-situ observation, predict human intention, quantify human distraction, and apply a non-stationary Markov Decision model to safely plan UAS motion in the presence of uncertainty.

\end{abstract}

\section{Introduction}
With advances in artificial intelligence, learning, and optimization approaches, copter-powered uncrewed aerial system (UAS) have found numerous applications in agriculture \cite{ahirwar2019application, mogili2018review}, warehouse industries \cite{wawrla2019applications, ali2023utilisation}, mining \cite{jones2019advances, jones2020applications}, surveillance \cite{dilshad2020applications, mishra2020drone}, construction \cite{umar2021applications, namian2021revealing}, and many other fields. Despite the widespread UAS applications, to date, UAS operation is closely supervised by human professionals. This paper, on the other hand, studies semi-autonomous missions that are jointly operated by UAS and human in the same environment. The primary focus of the paper is on ensuring safety of UAS operation in shared workplaces where humans and UAS coexist. Towards this goal, we combine Petri Nets with non-stationary Markov Decision Process (MDP) to provide a safe human-UAS collaboration abstraction model.

\subsection{Related Work}
Application of drones as safety inspector in a construction site was studied in Reference (Ref.) \cite{irizarry2012usability}. In Ref. \cite{cauchard2016emotion}, the authors rely on user experiences to incorporate human emotions into drone's trajectory planning. Refs. \cite{9151559, 9499075} study potential human-drone communication interfaces in a collaborative environment. Signal temporal logic \cite{raman2014model, deshmukh2017robust}  is used in Ref. \cite{10156559} to plan ergonomic and safe human-UAS collaboration in a construction environment. Researchers have used Ptri Nets to optimally schedule tasks to human and robot cooperating in a shared workplace \cite{8809746, casalino2019task, alirezazadeh2022dynamic, yagoda2012work, hjorth2022human}. Authors in  \cite{ali2024modeling} used colored Petri Nets (CPN)  for validation and verification of safety-critical systems \cite{ali2024modeling}. CPN was also used  in \cite{9636428} to model human actions, in a 
human-robot collaboration, under partial observability assumption.

\subsection{Contributions}
In the literature, researchers have extensively investigated autonomous UAS operation applications or UAS operation supervised by human professionals. This paper, on the other hand, aims to develop a model for semi-autonomous operations that are jointly conducted by UAS and human in a shared workplace. By considering human and UAS as co-workers, sharing a work environment, safety assurance becomes highly important since human and UAS co-workers can closely interact.  Because UAS need to deal with the uncertainty associated with incomplete knowledge about intention of each human co-worker, safety assurance is indeed very challenging. 

This paper proposes to use Petri Nets to model task evolution in a semi-autonomous operation, conducted by human and UAS co-workers. Assuming multiple Work Stations (WSs) exist, in a shared 2 workplace, the Petri Nets’ places abstractly represent WSs where the Petri Nets’s transitions can be used to either specify evolution of tasks in the same WS, or displacement between the WSs. Particularly, we use ``cyclic’’ construct to model evolution of incomplete task, or progression of multiple tasks, in the same place, when change of a WS is not needed. On the other hand, we use ``sequential,’’ ``dependency,’’ and ``conflict,’’to model displacement of UAS, or human, between WSs.

By applying the proposed model for abstraction of human-UAS collaboration, the paper offers the following novel contributions:

\begin{enumerate}
\item We enable UAS with on-board perception capability so that it can learn human intention and quantify human co-worker distraction by real-time processing of in-situ observations. By predicting human co-worker intention, UAS can resolve ``conflict’’ situations resulted from incomplete knowledge about human intentions. Also, UAS can quantify distraction of every human co-worker by specifying a probability distribution over a moving neighboring set (MNS). MNS is defined as a finite set of cells that form a rigid rectangular zone around each human coworker's desired trajectory. The MNS translational speed in the workplace is obtained using available search methods, such as A* search \cite{rastgoftar2022integration}.
\item We develop a non-stationary Markov Decision Process model for UAS motion planning in the presence of multiple non-cooperative human co-workers. While the state space, transition function, discount factor, and action components of the proposed MDP are time-invariant, the MDP cost is time-varying, and it is consistently updated so that learning of human intention and co-worker distraction are properly incorporated in the UAS motion planning.
\end{enumerate} 



\subsection{Outline}
This paper is organized as follows: Preliminary notions  of Petri Nets are reviewed in Section \ref{Preliminaries}. The problem of UAS-Human collaboration is formulated in Section \ref{Problem Formulation} and followed by the paper's approach for UAS onboard perception and motion planning in Section \ref{Approach}. Simulation results are presented in Section \ref{simulation}. Concluding remarks are stated in Section \ref{conclusion}.
\section{Preliminaries}\label{Preliminaries}
We use Petri Nets to model UAS-human collaboration in a constrained workplace.  The proposed  Petri Nets is defined by tuple $\mathcal{PN}=\left(\mathcal{P},\mathcal{T},\mathcal{E}_H,\mathcal{E}_U,\mathcal{W}_H,\mathcal{W}_U,\mathcal{M}_H,\mathcal{M}_U\right)$, where $\mathcal{P}$ is a finite set of places;  $\mathcal{T}$ is a finite set of transitions; $\mathcal{E}_H\subset \left(\mathcal{P}\times \mathcal{T}\right)\bigcup \left(\mathcal{T}\times \mathcal{P}\right)$ and $\mathcal{E}_U\subset \left(\mathcal{P}\times \mathcal{T}\right)\bigcup \left(\mathcal{T}\times \mathcal{P}\right)$  define  unweighted arcs of human and UAS co-workers, respectively; $\mathcal{W}_H:\mathcal{E}_H\rightarrow \left\{1\right\}$ defines weights of human co-worker transitions; $\mathcal{W}_U:\mathcal{E}_U\rightarrow \left\{1\right\}$  defines weights of UAS co-worker transitions;  marking $\mathcal{M}_{H}: \mathcal{P}\rightarrow \mathbb{N}$ specifies distribution of human co-workers; and   marking $\mathcal{M}_{U}: \mathcal{P}\rightarrow \mathbb{N}$ specifies distribution of UAS co-workers.  
In this work, set $\mathcal{P}$ defines actual Work Stations (WSs). 

\begin{figure}
    \centering
    \includegraphics[width=\linewidth]{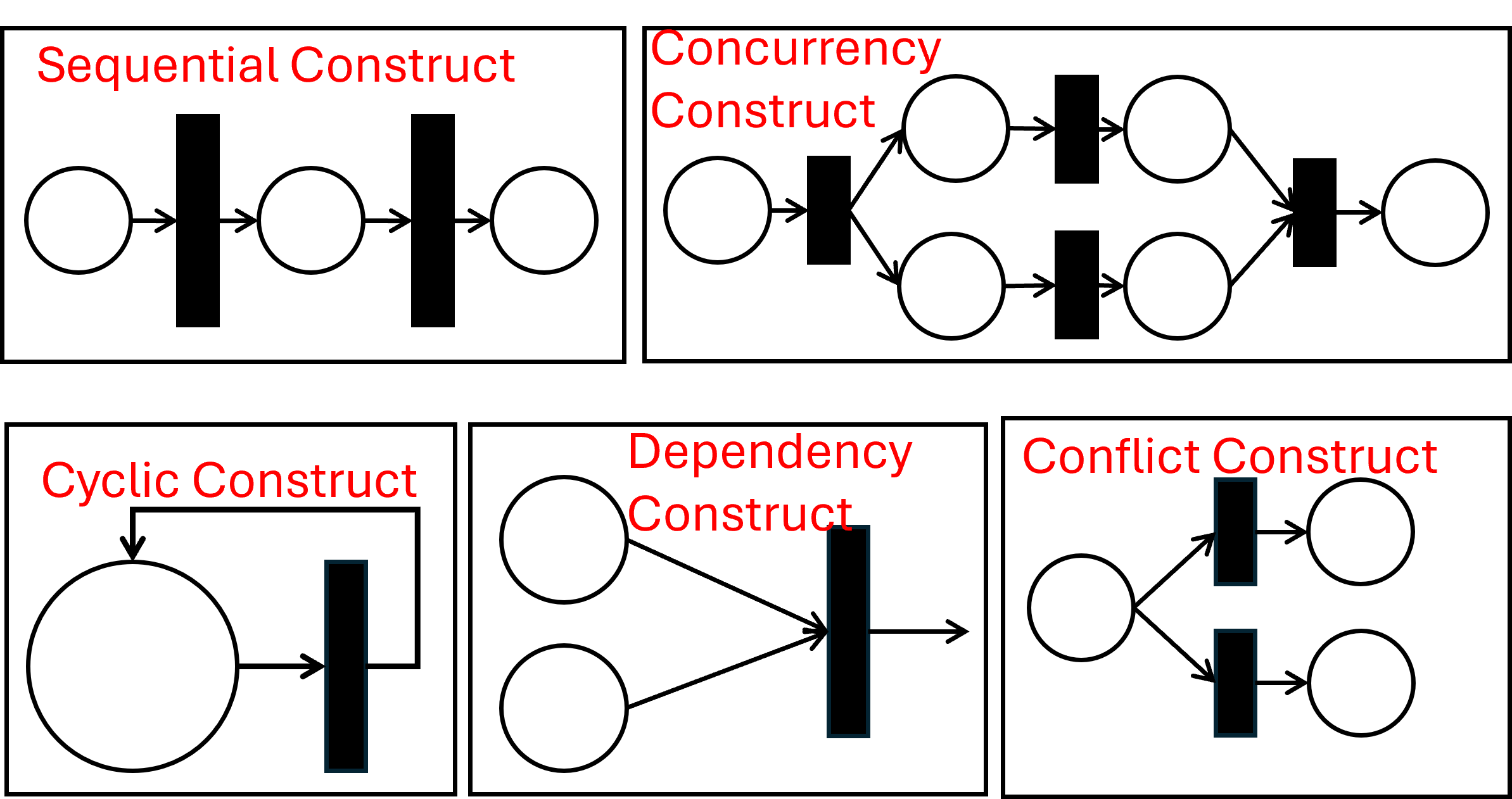}
    \caption{The constructs used for modeling human UAS collaboration.}
    \label{ContructsSchematic}
\end{figure}

\begin{definition}
    Given huaman arc set $\mathcal{E}_H$, we define out-neighbor transition set 
\begin{equation}
    \mathcal{N}_p^H=\left\{t\in \mathcal{T}:(p,t)\in \mathcal{E}_H\right\}
\end{equation}
for every place $p\in \mathcal{P}$.
\end{definition}
\begin{definition}
    Given UAS arc set $\mathcal{E}_U$, we define out-neighbor transition set 
\begin{equation}
    \mathcal{N}_p^U=\left\{t\in \mathcal{T}:(p,t)\in \mathcal{E}_U\right\}
\end{equation}
for every place $p\in \mathcal{P}$.
\end{definition}
\begin{definition}
    Set 
    \begin{equation}\label{epsilonprime}
    \mathcal{E}'=\left\{\left(p,t\right)\cup \left(t,p\right):\forall p\in \mathcal{P},~\forall t\in \mathcal{T}\right\}
    \end{equation}
    defines cyclic arcs (See the cyclic arc schematic in the bottom-left picture of Fig. \ref{ContructsSchematic}). 
\end{definition}
\begin{definition}
    Set 
    \begin{equation}
        \mathcal{R}_p^H=\left\{p'\in \mathcal{P}:\left(t,p'\right)\in \mathcal{E}_H\setminus \left(\mathcal{E}'\cap\mathcal{E}_H\right),~t\in \mathcal{N}_p^H\right\},\qquad p\in \mathcal{P}
    \end{equation}
    defines all possible  next WSs for \textit{human} co-workers.
\end{definition}
\begin{definition}
    Set 
    \begin{equation}
        \mathcal{R}_p^U=\left\{p'\in \mathcal{P}:\left(t,p'\right)\in \mathcal{E}_U\setminus \left(\mathcal{E}'\cap\mathcal{E}_U\right),~t\in \mathcal{N}_p^U\right\},\qquad p\in \mathcal{P}
    \end{equation}
    defines all possible  next WSs for \textit{UAS} co-workers.
\end{definition}
We note that $\mathcal{R}_p^H=\emptyset$, if a human co-worker does not change its current WS at $p\in \mathcal{P}$. Similarly, $\mathcal{R}_p^U=\emptyset$, if a UAS co-worker does not change its current WS at $p\in \mathcal{P}$.   

The Petri Nets, used for abstraction of human UAS collaboration, consists of five possible constructs that are shown in Fig. \ref{ContructsSchematic}. Functionality of these constructs  are described below:

\textbf{Cyclic Construct:} Proceeding of tasks in the same place $p\in \mathcal{P}$ is shown by the cyclic construct. Set $\mathcal{E}'$, given by Eq. \eqref{epsilonprime}, defines the non-weighted arcs that represent progression of tasks in the same WS. 

\textbf{Sequential Construct:} The sequential action is shown by the top-left picture  in Fig.  \ref{ContructsSchematic}. Note that the sequential action represents the situations where the transition of the  human co-worker  is known because the worker has only one possible action to follow next. Also, sequential constructs are used to specify UAS displacement between WSs.

\textbf{Conflict Construct:} The conflict construct  is used to incorporate UAS uncertainty associated with the human decision/intention for choosing the next place $p'\in \mathcal{R}_p^H$. For place $p\in \mathcal{P}$ dealing with a conflict construct, $\left|\mathcal{R}_p^H\right|>1$.

\textbf{Dependency Construct:} The dependency construct 
represents a situation at which firing a transition $t\in \mathcal{T}$ requires existence of token in multiple input places. 

\textbf{Concurrency Construct:} The concurrency construct shown in Fig. \ref{ContructsSchematic} represents a situation at which a human co-worker, or UAS, is conducting multiple tasks simultaneously. 

    

We define the following rules for activation of the arcs:

\textbf{Rule 1:} The markings $\mathcal{M}_H(p)\geq 0$ and $\mathcal{M}_U(p)\geq 0$ at every place  $p\in \mathcal{P}$. If $M_H(p)=0$, there is no human co-worker in place $p\in \mathcal{P}$. Otherwise, $M_H(p)>0$ assigns the number of human co-workers in place $p\in \mathcal{P}$. Similarly, $\mathcal{M}_U(p)\geq 0$ specifies the number of UAS co-worker in place $p\in \mathcal{P}$.

\textbf{Rule 2:} A transition $t$ is fired for a human co-worker at $p\in \mathcal{P}$, if the following  
\begin{equation}
    \bigwedge_{p\in \mathcal{P}}\bigwedge_{t\in \mathcal{T}}\bigwedge_{\left(p,t\right)\in \mathcal{E}_H}\left(\mathcal{M}_H(p)-\mathcal{W}_H(p,t)\geq0\right),
\end{equation}
where $\bigwedge$ means ``include all''.

\textbf{Rule 3:} A transition $t$ is fired for a UAS co-worker at $p\in \mathcal{P}$, if the following  
\begin{equation}
    \bigwedge_{p\in \mathcal{P}}\bigwedge_{t\in \mathcal{T}}\bigwedge_{\left(p,t\right)\in \mathcal{E}_U}\left(\mathcal{M}_U(p)-\mathcal{W}_U(p,t)\geq0\right),
\end{equation}
where $\bigwedge$ means ``include all''.

\textbf{Rule 4:}  To ensure safety of the human co-workers, UAS co-workers can be present at WS $p\in \mathcal{P}$, if the following condition is met:
\begin{equation}\label{rule4}
\mathcal{M}_U(p)+\mathcal{M}_H(p)=
    \begin{cases}
        \leq2& \mathcal{M}_U(p)>0\\
        \geq 0&\mathcal{M}_U(p)=0
    \end{cases}
    ,\qquad \forall p\in \mathcal{P}.
\end{equation}
By imposing condition \eqref{rule4}, the following safety regulations are imposed for the presence of  UAS and human  co-workers is WS $p\in \mathcal{P}$:
\\
\textit{Regulation 1:} No constraints are imposed on the number of human co-workers in WS $p\in \mathcal{P}$, if no UAS exists in $p\in \mathcal{P}$.
\\
\textit{Regulation 2:} The number of UAS and human co-workers cannot exceed $2$ in WS $p\in \mathcal{P}$, if one UAS exists in WS $p\in \mathcal{P}$.
\\
\textit{Regulation 2:} At most  $2$ UAS co-worker can work in WS $p\in \mathcal{P}$, if no human coworker is present in WS $p\in \mathcal{P}$.




\section{Problem Formulation}\label{Problem Formulation}
We consider operation of UAS in the presence of $N_H$ human co-workers identified by set $\mathcal{H}=\left\{1,\cdots,N_H\right\}$ in a shared workplace. The entire workplace is spatially discretized and  abstractly represented by finite set $\mathcal{D}=\mathcal{X}\times \mathcal{Y}$ where $\times$ is the Cartesian product symbol;  $\mathcal{X}=\left\{1,\cdots,n_x\right\}$ and $\mathcal{Y}=\left\{1,\cdots,n_y\right\}$ identify the $x$ and $y$ coordinates of the cells, respectively. We also define $\mathcal{O}\subset \mathcal{D}$ as stationary obstacles in the workplace that cannot be reached neither by UAS or a human co-worker. The WS  positions are defined by finite set 
\begin{equation}
    \bar{\mathcal{D}}=\left\{\mathbf{r}_i=\left(\bar{x}_i,\bar{y}_i\right)\in \mathcal{D}\setminus \mathcal{O}:i\in \mathcal{P}\right\}.
\end{equation}
We consider scenarios at which a UAS co-worker aims to reach $j\in \mathcal{R}_i^U$ from  $i\in \mathcal{P}$, where the UAS motion can be abstracted by a sequential construct through firing transition $t_u\in \mathcal{T}$. Therefore, $\left(i,t_u\right)\bigcup\left(t_u,j\right)\in \mathcal{E}_U$. On the other hand, the human co-worker decision is unknown to UAS, therefore, the UAS uses a conflict construct to model  human co-worker behavior.

Given above problem setting, the paper studies the following three problems:

\textbf{Problem 1: Human Distraction Quantification:} The first problem aims to quantify the \textit{distraction} of human co-worker $h\in \mathcal{H}$  moving from $p_h\in \mathcal{P}$ to $p'_h\in \mathcal{R}_{p_h}^H$ ($p'_h\neq p_h$). To this end, we first specify an MNS around the desired trajectory of human co-worker $h\in \mathcal{H}$ which is  obtained by using available search method. We then quantify human distraction through real-time processing of in-situ observations of human motion.

\textbf{Problem 2: Human Intention Prediction:} The second problem is to learn intention of human co-worker $h\in \mathcal{H}$ by quantifying the intention probability $\mathrm{Pr}(p'_h|p_h)$ for $p_h\in \mathcal{P}$ and $p'_h\in \mathcal{R}_{p_h}^H$, where 
\begin{equation}
    \sum_{p'_h\in \mathcal{R}_{p_h}^H}\mathrm{Pr}(p'_h|p_h)=1,\qquad p_h\in \mathcal{P},~h\in \mathcal{H}.
\end{equation}

\textbf{Problem 3: UAS Motion Planning:} The third problem is to obtain a safe trajectory for a  UAS in a workplace shared with human co-workers. The key assumption of the paper is that UAS flies at low altitude close to the ground. This assumption is necessary for indoor UAS operation in some workplaces such as  underground mines, warehouse sites, or under-canopy spaces in farms. Therefore, the objective of trajectory planning is that  UAS avoids areas where the likelihood of existence of human co-worker is high.
We develop a non-stationary MDP to optimize the UAS trajectory.

\section{Approach}\label{Approach}


Problems 1, 2, and 3 are formulated and explained in Sections \ref{Problem 1}, \ref{Problem 2}, and \ref{Problem 3} below. We let $p_h(k)\in \mathcal{P}$ and $p'_h(k)\in \mathcal{R}_p^H$ be the ``origin'' WS and ``destination'' WS place of a  human co-worker  $h\in \mathcal{H}$ at discrete time $k$. We use the following position notations:
\begin{enumerate}
    \item $\bar{\mathbf{r}}_{h,p_h,p'_h}\in \bar{\mathcal{D}}$: ``Current'' desired position of human co-worker $h\in \mathcal{H}$ given $p_h(k)\in \mathcal{P}$ and $p'_h(k)\in \mathcal{R}_{p_h}^H$ as the ``origin'' WS and ``destination'' WS place, respectively, at discrete time $k$.
    \item $\bar{\mathbf{r}}_{h,p_h,p'_h}^+\in \bar{\mathcal{D}}$: ``Next'' desired position of human co-worker $h\in \mathcal{H}$ given $p_h(k)\in \mathcal{P}$ and $p'_h(k)\in \mathcal{R}_{p_h}^H$ as the ``origin'' WS and ``destination'' WS place, respectively,  at discrete time $k$.
    \item $\mathbf{r}_{h}$: ``Actual'' position of human worker $h\in \mathcal{H}$ at discrete time $k$.
    \item $\mathbf{r}_U$: ``Actual'' position of UAS at discrete time $k$.
\end{enumerate}
Note that the desired trajectory of human worker $h\in \mathcal{H}$ can be assigned by using an A* search over set $\mathcal{D}$. Also, actual position of every human co-worker $h\in \mathcal{H}$ is known to the UAS at every discrete time $k$, where it is either captured by on-site cameras and communicate to the UAS, or it is accurately estimated by the UAS through processing onboard visual sensory information.
\subsection{Problem 1: Distraction  Probability Quantification}\label{Problem 1}

    We denote the  desired velocity of human worker $h\in \mathcal{H}$ by $\bar{\mathbf{v}}_h$ and define it as $\bar{\mathbf{v}}_h=\bar{\mathbf{r}}_{h,p_h,p'_h}^+-\bar{\mathbf{r}}_{h,p_h,p'_h}$. Given $\bar{\mathbf{v}}_h$, the motion of human worker $h\in \mathcal{H}$ is categorized as follows:
\begin{itemize}
    \item We say  human co-worker $h\in \mathcal{H}$ moves ``diagonally,'' if $\|\bar{\mathbf{v}}_h\|=\sqrt{2}$. 
    \item We say  human co-worker $h\in \mathcal{H}$ moves ``straight,'' if $\|\bar{\mathbf{v}}_h\|=1$.  
    \item A human co-worker $h\in \mathcal{H}$ desires to ``Stay'',  if $\bar{\mathbf{v}}_h=\left(0,0\right)$.
\end{itemize}
\begin{definition}
For every human worker $h\in \mathcal{H}$, 
\begin{equation}
 \begin{split}
    \mathcal{I}_d\left(\bar{\mathbf{r}}_{h,p_h,p'_h}\right)=&\bigg\{\bar{\mathbf{r}}_{h,p_h,p'_h}+\left(i_s,j_s\right)\in \mathcal{D}:\\
    &\left(i_s,j_s\in \left\{0,\cdots,d\right\}\right)\wedge\left(\left(i_s\neq 0\right)\vee \left(j_s\neq 0\right)\right)\bigg\}.
\end{split}
\end{equation}
is the moving neighboring set (MNS) of degree $d$, where $d$ is called \textit{degree of neighborhood}. 
\end{definition}
Because  $d$ is time-invariant, MNS remains rigid at every discrete time $k$. 
\begin{definition}
    We use $b_{h}\left(\mathbf{r}_h, \bar{\mathbf{r}}_{h,p_h,p'_h},\|\bar{\mathbf{v}}_h\|\right)$ to denote  the number of visit of  $\mathbf{r}_h\in \mathcal{I}_d\left(\bar{\mathbf{r}}_{h,p_h,p'_h}\right)$, when human co-worker $h\in \mathcal{H}$ moves from origin $p_h(k)\in \mathcal{P}$ towards destination $p'_h(k)\in \mathcal{R}_{p_h}^H$ with desired velocity $\bar{\mathbf{v}}_h$.
\end{definition}


The deviation probability of human worker $h\in \mathcal{H}$ is denoted by $\alpha_h$ and defined by
\begin{equation}\label{distractionprobability}    \alpha_h\left(\mathbf{r}_h|\bar{\mathbf{r}}_{h,p_h,p'_h},\|\bar{\mathbf{v}}_h\|\right)=\dfrac{b_h\left(\mathbf{r}_h,\bar{\mathbf{r}}_{h,p_h,p'_h},\|\bar{\mathbf{v}}_h\|\right)}{\sum_{\mathbf{r}_h\in \mathcal{I}_d\left(\bar{\mathbf{r}}_{h,p_h,p'_h}\right)}b_h\left(\mathbf{r}_h,\bar{\mathbf{r}}_{h,p_h,p'_h},\|\bar{\mathbf{v}}_h\|\right)}
\end{equation}
for every human worker $h\in \mathcal{H}$. 
\subsection{Problem 2: Human Intention Prediction}\label{Problem 2}
We define 
\begin{equation}
    \delta\left(\mathbf{r}_h(k),\bar{\mathbf{r}}_{h,p_h,p'_h}\left(k\right)\right)=\mathrm{exp}\left(-\left\|\mathbf{r}_h(k)-\bar{\mathbf{r}}_{h,p_h,p'_h}(k)\right\|\right),
\end{equation}
to quantify deviation of human co-worker $h\in \mathcal{H}$ from the desired trajectory $\bar{\mathbf{r}}_{h,p_h,p'_h}$ when the worker moves from origin $p_h\in \mathcal{P}$, towards destination $p'_h\in \mathcal{R}_{p_h}^H$. Note that   $\delta\left(\mathbf{r}_h,\bar{\mathbf{r}}_{h,p_h,p'_h}\right)$ is a reward function that decreases from $1$ to $0$ as $\left\|\mathbf{r}-\bar{\mathbf{r}}_{h,p_h,p'_h}\left(p,p'\right)\right\|$ increases from $0$ to $+\infty$.  

By knowing $p_h\in \mathcal{P}$, $p'_h\in \mathcal{R}_{p_h}^H$,  trajectory of human worker $h\in \mathcal{H}$, over the past $N_p$ time steps, 
\begin{equation}\label{intentionprobability}
    \mathrm{Pr}\left(p'_h|p_h\right)={\sum_{\tau=k-N_p}^{k-1}\delta\left(\mathbf{r}_h(\tau),\bar{\mathbf{r}}_{h,p_h,p'_h}\left(\tau\right)\right)\over \sum_{p'_h\in \mathcal{R}_p^H}\sum_{\tau=k-N_p}^{k-1}\delta\left(\mathbf{r}_h(\tau),\bar{\mathbf{r}}_{h,p_h,p'_h}\left(\tau\right)\right)}
\end{equation}
assigns the probability that $p'_h\in \mathcal{R}_{p_h}^H$ is the goal WS for human worker $h\in \mathcal{H}$ at discrete time $k$.
\subsection{Problem 3: UAS Motion Planning}\label{Problem 3}
We consider a scenario at which a UAS is in transitioning mode from WS $i\in \mathcal{P}$ to goal WS $g\in \mathcal{R}_i^U$ while interacting with multiple human co-workers. The UAS motion planning is considered as a non-stationary Markov Decision Problem with the components and solution presented in Sections \ref{MDP Components} and \ref{Solution}.

\subsubsection{MDP Components}\label{MDP Components}
The proposed (non-stationary) MDP is given by tuple $\left(\mathcal{S},\mathcal{A},\mathcal{C}_{g,k},\mathcal{F},\gamma,g\right)$ with state set $\mathcal{S}$, action set $\mathcal{A}$, transition function $\mathcal{F}:\mathcal{S}\times \mathcal{A}\rightarrow \mathcal{S}$, cost $\mathcal{C}_{g,k}:\mathcal{S}\times \mathbb{N}\rightarrow \mathbb{R}$, discount factor $\gamma\in \left[0,1\right]$, and UAS goal WS $g\in \mathcal{P}$ that is positioned at $\mathbf{r}_g\in \bar{\mathcal{D}}$. These components are defined as follows:

\textit{\underline{State Set $\mathcal{S}$:}} 
We use set  ${\mathcal{K}}=\left\{1,\cdots,N_{\tau}\right\}$ to define a future horizon interval of length $N_\tau$. Then, the MDP state space is defined by
\begin{equation}\label{StateSet}
\mathcal{S}=\left\{\mathbf{s}=\left(\mathbf{r},\tau\right):\mathbf{r}\in \mathcal{D},~\tau\in \mathcal{K}\right\}.
\end{equation}

\textit{\underline{Action Set $\mathcal{A}$:}} We  use set 
\begin{equation}
    \mathcal{A}=\left\{ \mathrm{E}, \mathrm{NE}, \mathrm{N}, \mathrm{NW}, \mathrm{W}, \mathrm{SW}, \mathrm{S}, \mathrm{SE}, \mathrm{O} \right\}
\end{equation}
to define nine possible actions including go toward ``east’’ (E), go toward ``northeast’’ (NE), go towards ``north’’ (N), go toward ``northwest’’ (NW), go towards ``west’’ (W), go towards ``southwest’’ (SW), go towards ``south’’ (S), go towards ``southeast’’ (SE), and ``no motion’’ (O).

\textit{\underline{Cost $\mathcal{C}_{g,k}$:}} The MDP cost depends only on state $\mathbf{s}\in \mathcal{S}$. By expressing $\mathbf{s}=\left(\mathbf{r},\tau\right)$, for every $\mathbf{r}\in \mathcal{D}$ and every $k\in \mathcal{K}$,   cost is imposed on operation of the UAS that aims to move from the ``origin'' WS $i\in \mathcal{P}$ towards ``goal'' WS $g\in \mathcal{R}_i^U$ at discrete. We note that the MDP cost does not depend on on the origin of the UAS WS biut it depends of the UAS goal WS $g\in \mathcal{P}$. To define the MDP cost, we define the spatiotemporal heat map
\begin{equation}    \mathcal{J}_{g,k}\left(\mathbf{s}\right)=\bar{\mathcal{J}}_g\left(\mathbf{r}\right)+\underline{\mathcal{J}}_k\left(\mathbf{r},\tau\right),
\end{equation}
where 
\begin{equation}
    \bar{\mathcal{J}}_g\left(\mathbf{r}\right)=\|\mathbf{r}-\mathbf{r}_g\|,\qquad i\in \mathcal{P},~g\in \mathcal{R}_i^U,~\mathbf{r},\mathbf{r}_g\in \mathcal{D},
\end{equation}
is the distance from $\mathbf{r}\in \mathcal{D}$ and $\mathbf{r}_j\in \mathcal{D}$ and
\begin{equation}
\resizebox{0.99\hsize}{!}{%
$
\begin{split}
    \underline{\mathcal{J}}_k\left(\mathbf{r},\tau\right)=&c_0\sum_{\tau\in \mathcal{K}}\sum_{h\in \mathcal{H}}\sum_{p_h\in \mathcal{P}}\sum_{p'_h\in \mathcal{R}_{p_h}^H}\mathrm{Pr}\left(p'_h(k+\tau)|p_h(k+\tau)\right)\times\\
    &\alpha_h\left(\mathbf{r}|\bar{\mathbf{r}}_{h,p_h,p'_h}(k+\tau ), \bar{\mathbf{v}}_{h}(k+\tau)\right)
\end{split}    
    $
    }
\end{equation}
where $c_0>0$, $\tau\in \mathcal{K}$, and $\mathbf{r}\in \mathcal{D}$. The MDP cost is then defined by
\begin{equation}\label{MDPCost}
    \mathcal{C}_{g,k}\left(\mathbf{r},\tau\right)=\begin{cases}
        c_g<<0&\mathbf{r}=\mathbf{r}_g\in \bar{\mathcal{D}},~g\in \mathcal{P},~\forall k,~\forall \tau\\
        c_{obs}>>0&\mathbf{r}\in \mathcal{O},~\forall k,~\forall \tau\\
        \mathcal{J}_{g,k}\left(\mathbf{s}\right)&\mathbf{r}\neq\mathbf{r}_g,~\mathbf{r}\notin \mathcal{O},~\forall k,~\forall \tau\\
    \end{cases}
    ,
\end{equation}
where  $\left|c_g\right|$ and  $\left|c_{obs}\right|$ are sufficiently large and constant.

\textit{\underline{Transition Probability:}} Transition probability function $\mathcal{F}$ specifies transitions of the UAS  over the state space under all possible MDP actions. By assuming that UAS actuators and sensors are all healthy, we can suppose deterministic transitions over the state space. Under this assumption, $\mathcal{F}\left(\mathbf{s}^+|\mathbf{s},a\right)\in \left\{0,1\right\}$ is a binary variable, for every current state $\mathbf{s}\in\mathcal{S}$, next state $\mathbf{s}^+\in\mathcal{S}$, and action $a\in \mathcal{A}$. We use Algorithm \ref{alg2} to specify deterministic transitions of the UAS over the state space $\mathcal{S}$.
\begin{algorithm}
  \caption{Obtaining function $\mathcal{F}\left(\mathbf{s}^+|\mathbf{s},a\right)$ under deterministic transition assumption.}\label{alg2}
  \begin{algorithmic}[1]
          \State \textit{Get:} Obstacle set $\mathcal{O}$; $n_x=\left|{\mathcal{X}}\right|$; $n_y=\left|{\mathcal{Y}}\right|$; $n_\tau=\left|{\mathcal{K}}\right|$; $a\in \mathcal{A}$.
         \State \textit{Return:} Binary function $\mathcal{F}\left(\mathbf{s}^+|\mathbf{s},a\right)$.
          \For{\texttt{ $k\in {\mathcal{K}}$}}
          \If{$k\leq  N_\tau$}           
               \State $k'\leftarrow k+1$.      
          \EndIf
              \For{\texttt{ $i\in {\mathcal{X}}$}}
                \For{\texttt{ $i\in {\mathcal{Y}}$}}
                    \State $\mathbf{s}\leftarrow \left(i,j,k\right)$.  
                    \State $i'\leftarrow i$ and  $j'\leftarrow j$.
                    \If{$i<n_x$ and $a=\left\{\mathrm{SE},\mathrm{E},\mathrm{NE}\right\}$} 
                        \State $i'\leftarrow i+1$. 
                    \EndIf
                    \If{$i>n_x$ and $a=\left\{\mathrm{SW},\mathrm{W},\mathrm{NW}\right\}$} 
                        \State $i'\leftarrow i-1$. 
                    \EndIf
                    \If{$j<n_y$ and $a=\left\{\mathrm{NE},\mathrm{N},\mathrm{NW}\right\}$} 
                        \State $j'\leftarrow j+1$. 
                    \EndIf
                    \If{$j>1$ and $a=\left\{\mathrm{SE},\mathrm{S},\mathrm{SW}\right\}$} 
                        \State $j'\leftarrow j-1$. 
                    \EndIf
                    \State $\mathbf{s}^+\leftarrow \left(i',j',k'\right)$.
                    \State $\mathcal{F}\left(\mathbf{s}^+|\mathbf{s},a\right)\leftarrow 1$.
                \EndFor
              \EndFor
        \EndFor  
  \end{algorithmic}
\end{algorithm}
\begin{figure}
    \centering
    \includegraphics[width=\linewidth]{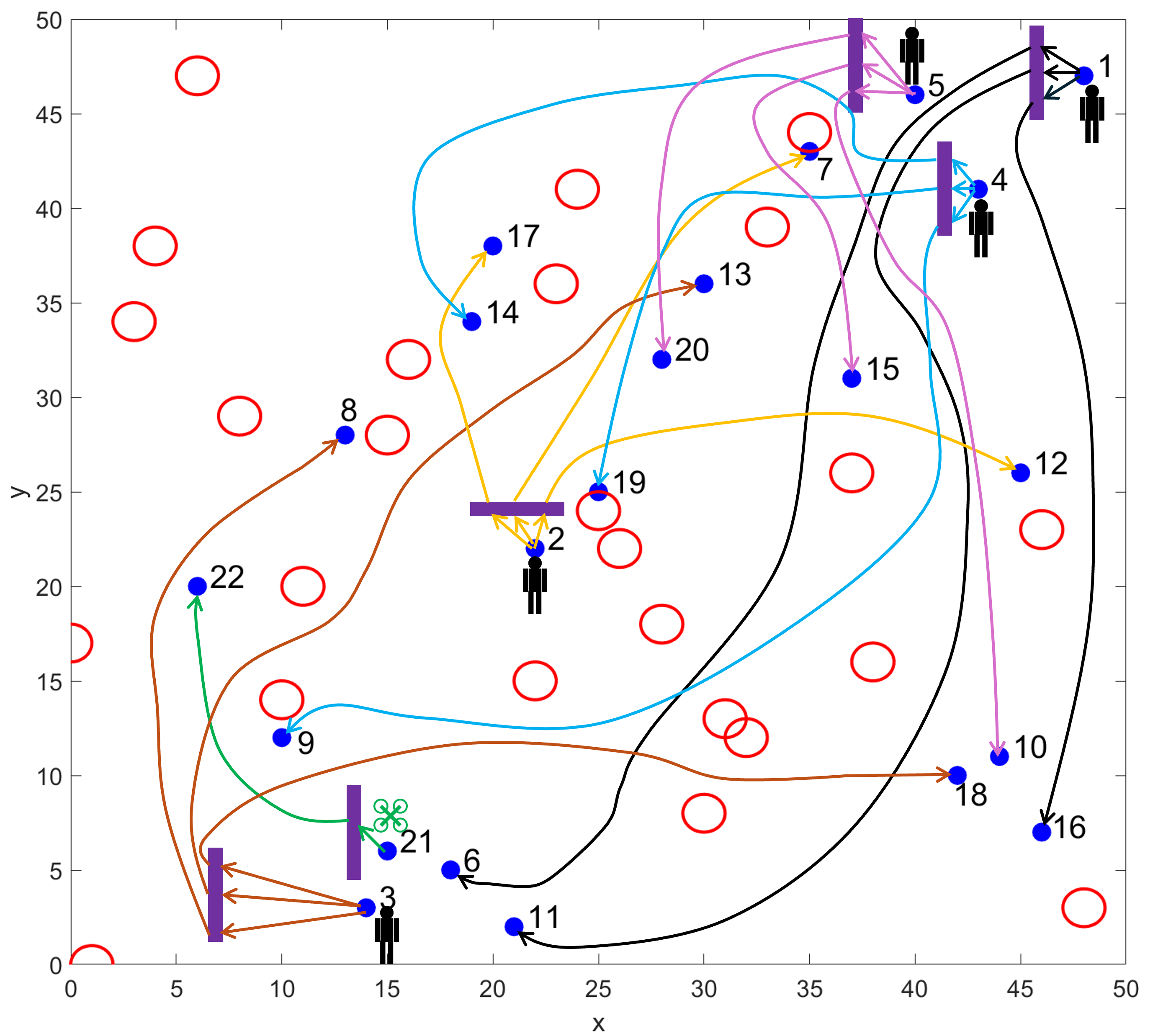}
    \caption{The Petri Nets demonstrating human UAS interaction in a shared workplace.}
    \label{PetrinetsimulationWithFrame}
\end{figure}
\begin{figure}[h]
\centering
\subfigure[]{\includegraphics[width=0.49\linewidth]{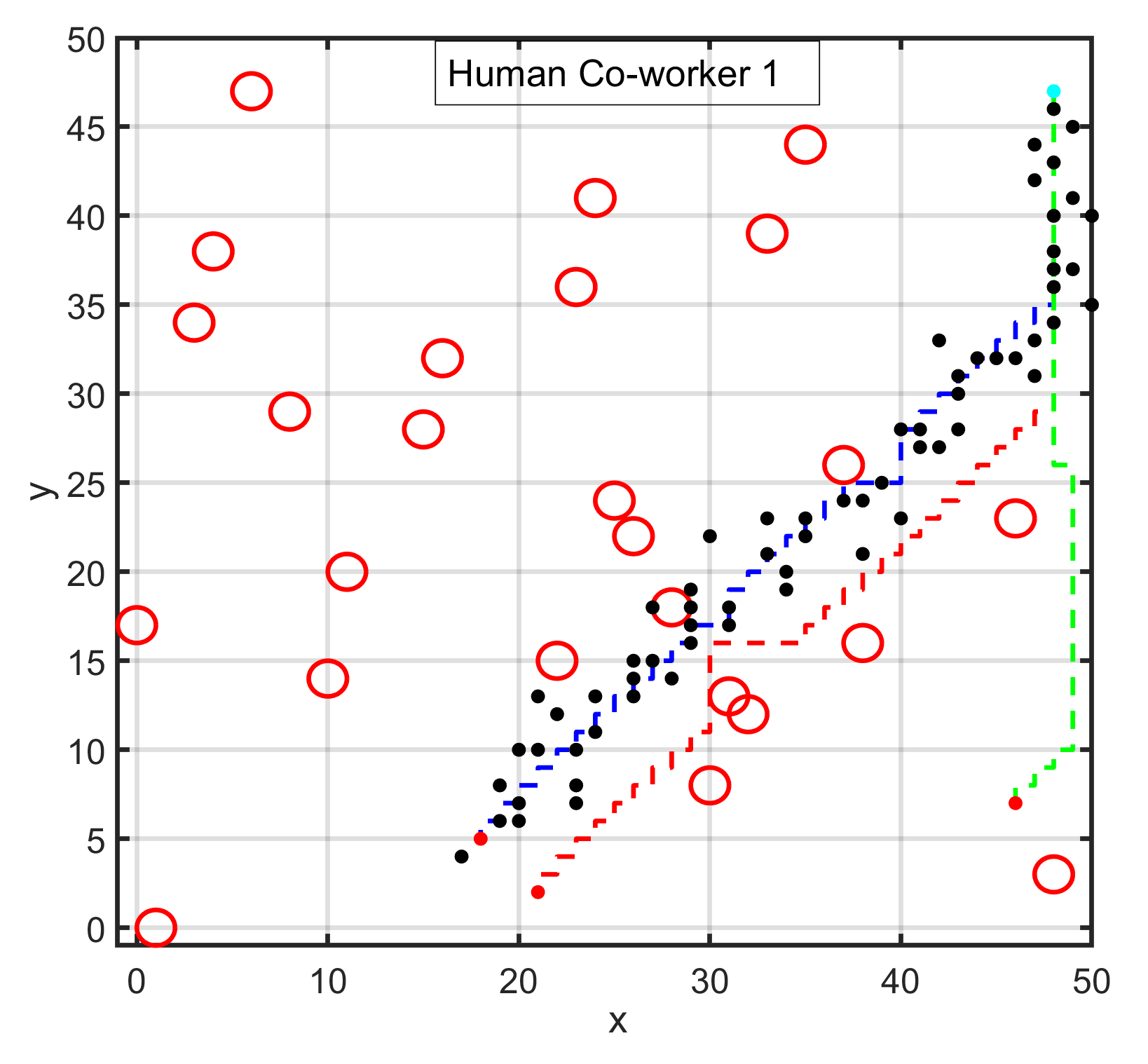}}
\subfigure[]{\includegraphics[width=0.49\linewidth]{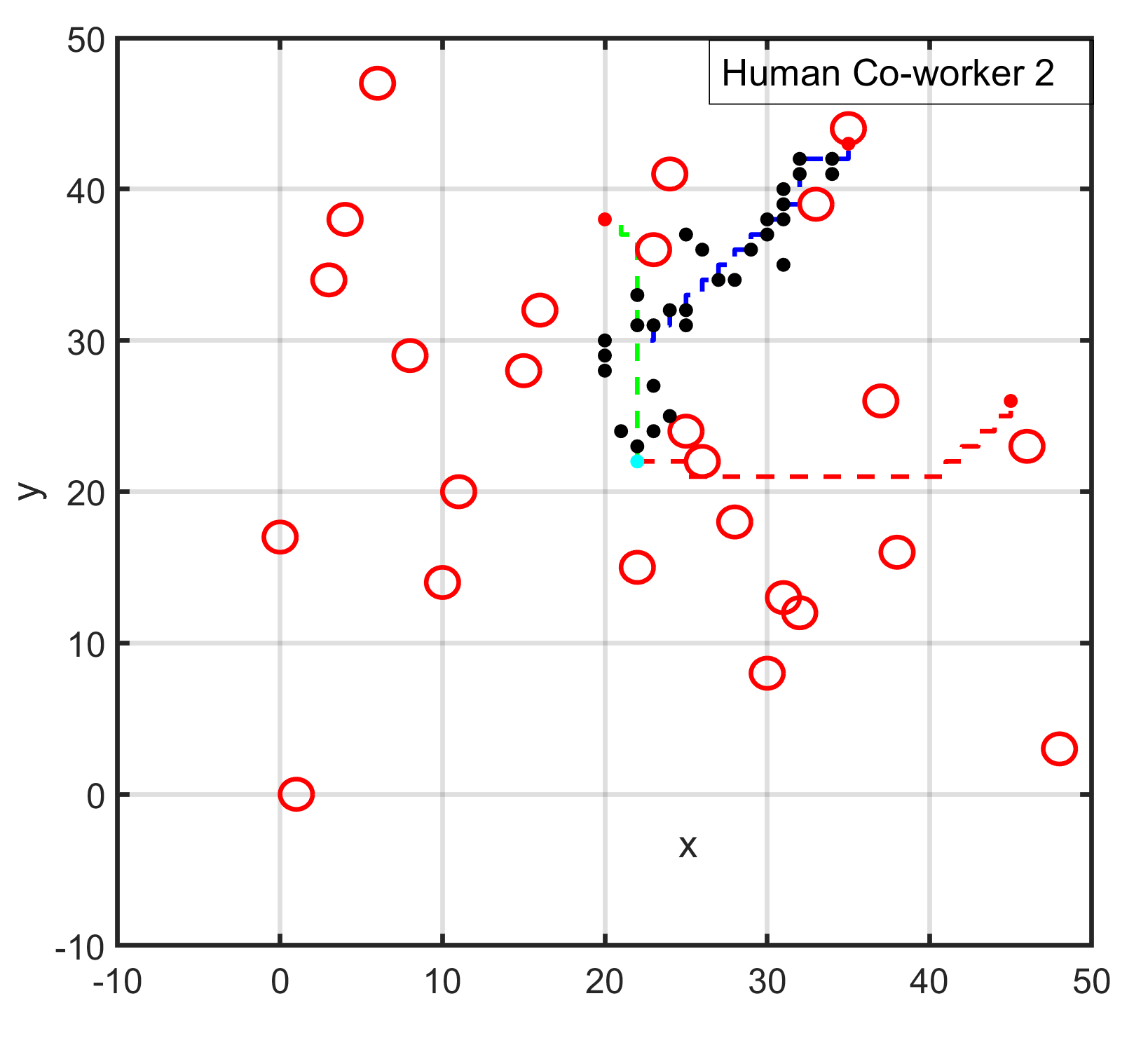}}
\subfigure[]{\includegraphics[width=0.49\linewidth]{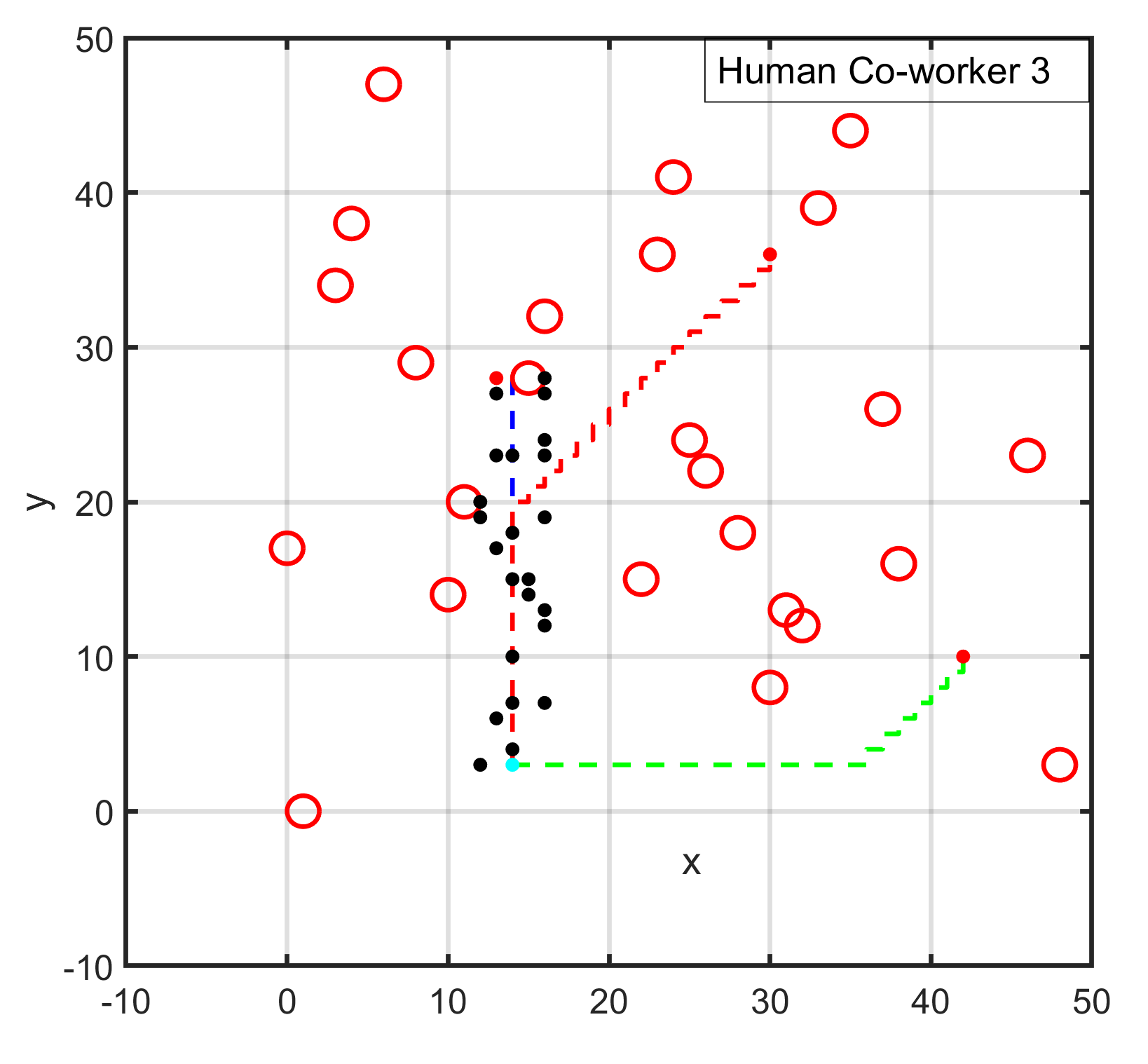}}
\subfigure[]{\includegraphics[width=0.49\linewidth]{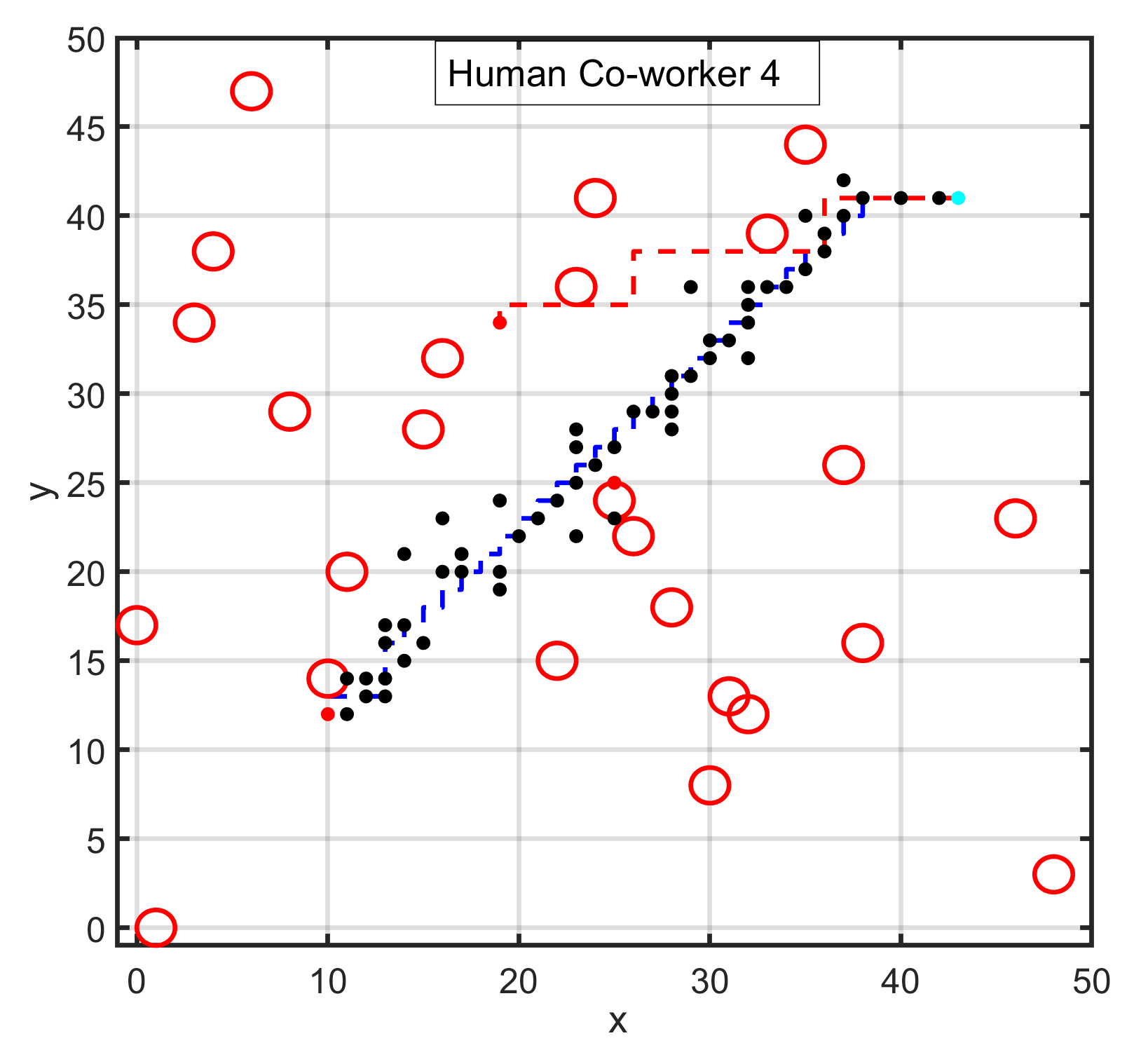}}
\subfigure[]{\includegraphics[width=0.49\linewidth]{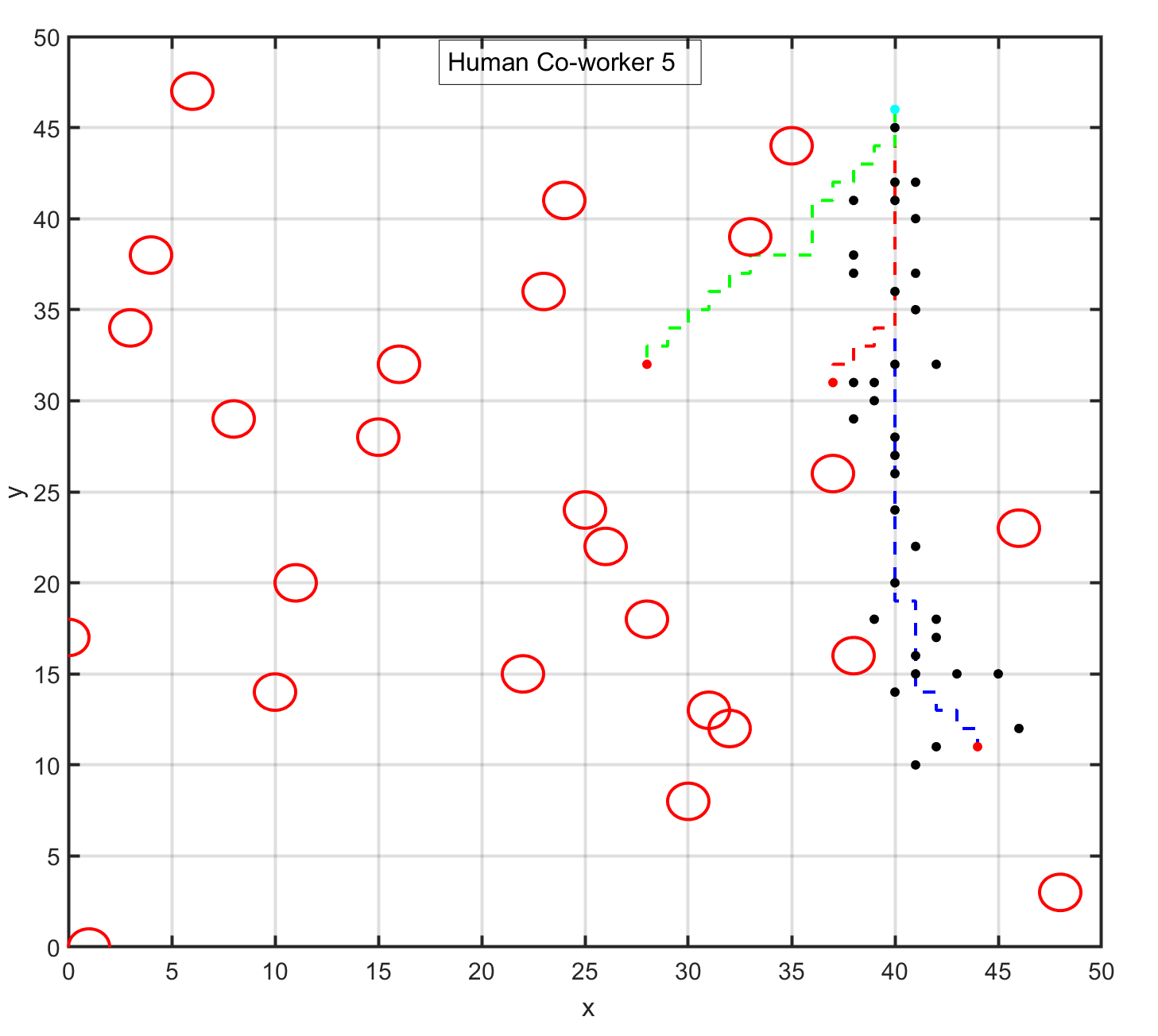}}
\caption{Desired trajectories between the origin WS and  three possible next WS are shown by dashed plot for human co-workers $1$, $2$, $3$, $4$, and $5$ in sub-figures (a), (b), (c), (d), and (e), respectively. Also, actual trajectory of each co-worker is shown by black dots in sub-figures (a)-(e).}
\label{Humanco-worker1Paths}
\end{figure}
\begin{figure}[h]
\centering
\subfigure[Human co-worker 1]{\includegraphics[width=0.98\linewidth]{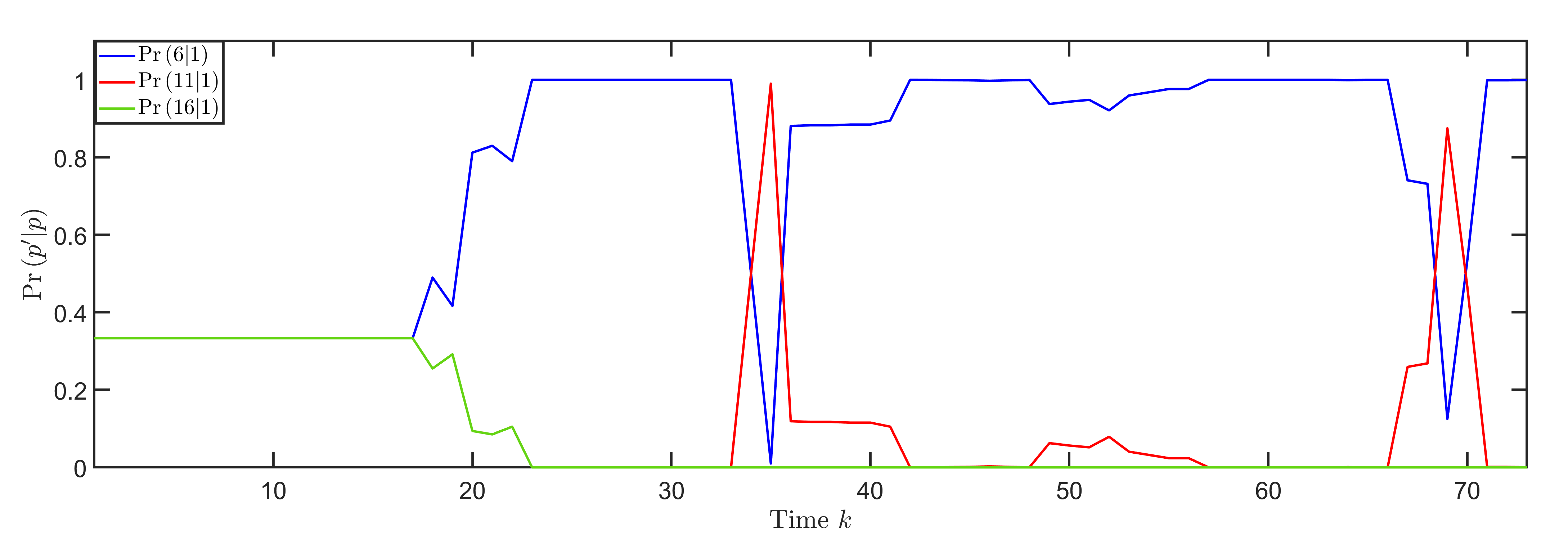}}
\subfigure[Human co-worker 2]{\includegraphics[width=0.98\linewidth]{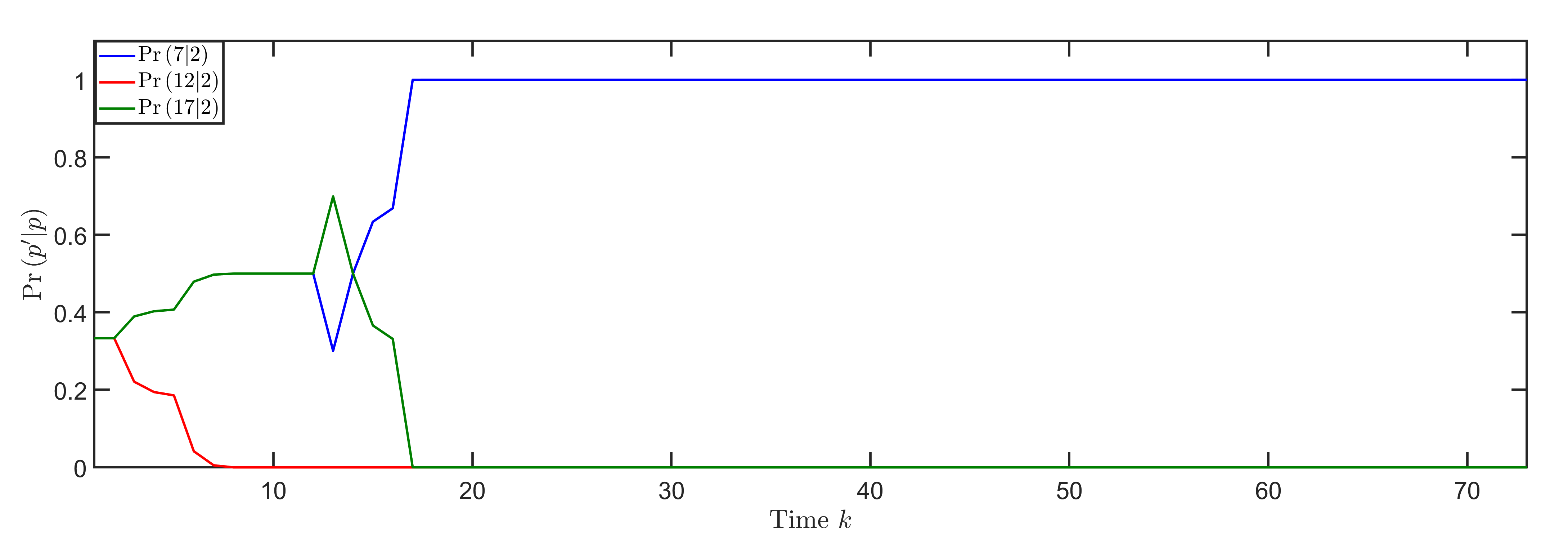}}
\subfigure[Human co-worker 3]{\includegraphics[width=0.98\linewidth]{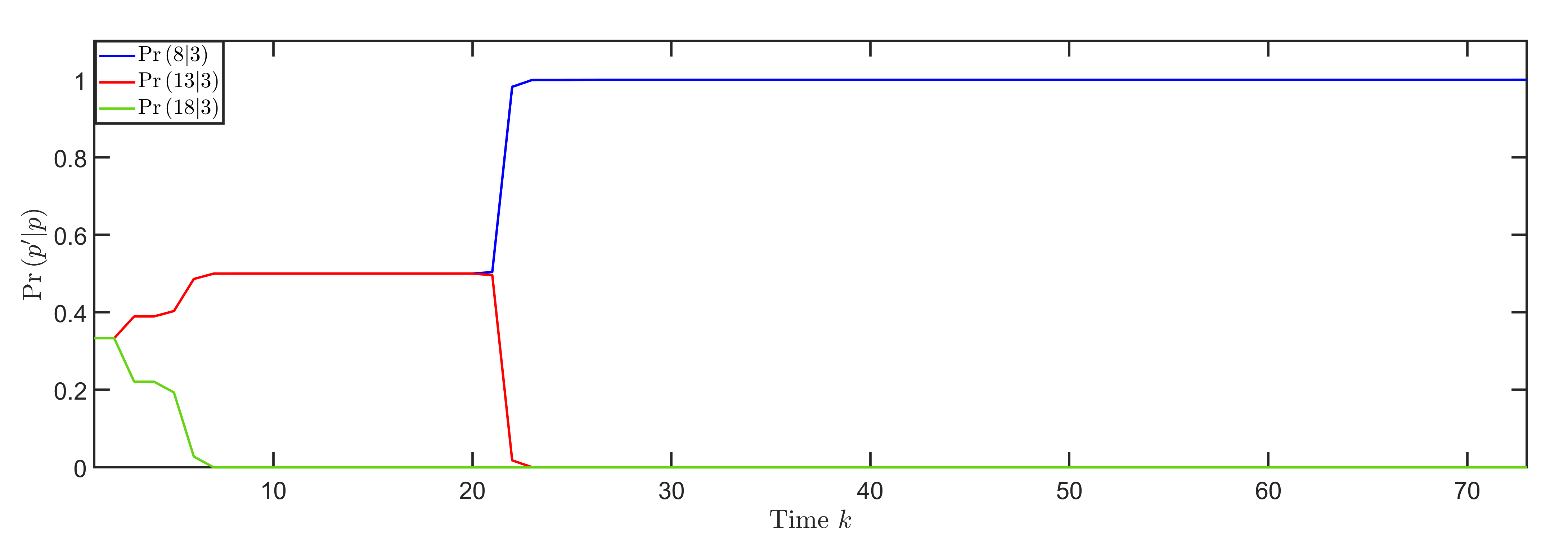}}
\subfigure[Human co-worker 4]{\includegraphics[width=0.98\linewidth]{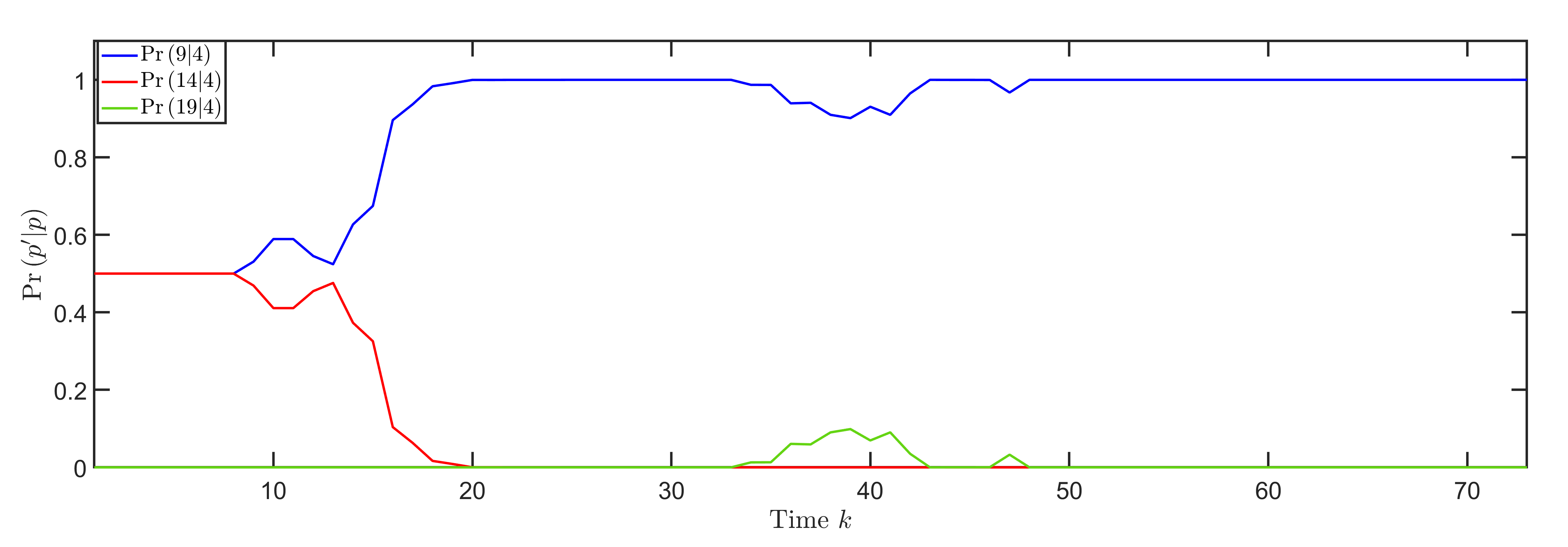}}
\subfigure[Human co-worker 5]{\includegraphics[width=0.98\linewidth]{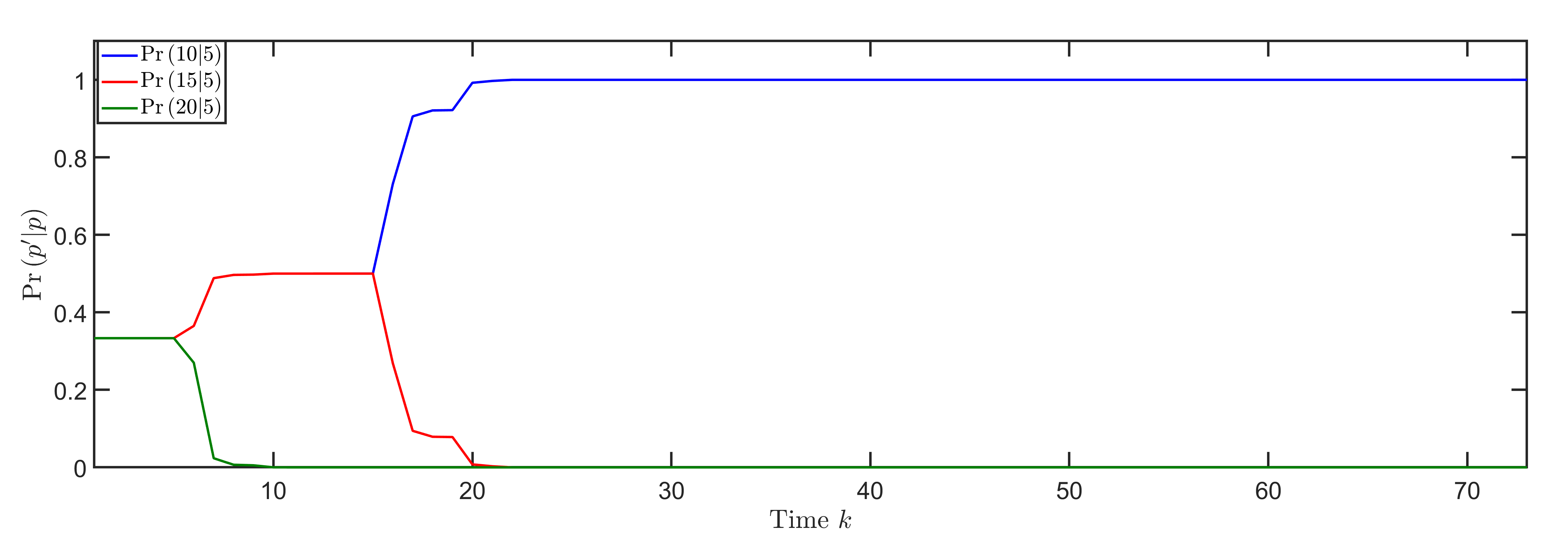}}
\caption{Intention prediction of human co-workers.}
\label{IntentionProbabilityHuman}
\end{figure}
\begin{figure}
    \centering
    \includegraphics[width=\linewidth]{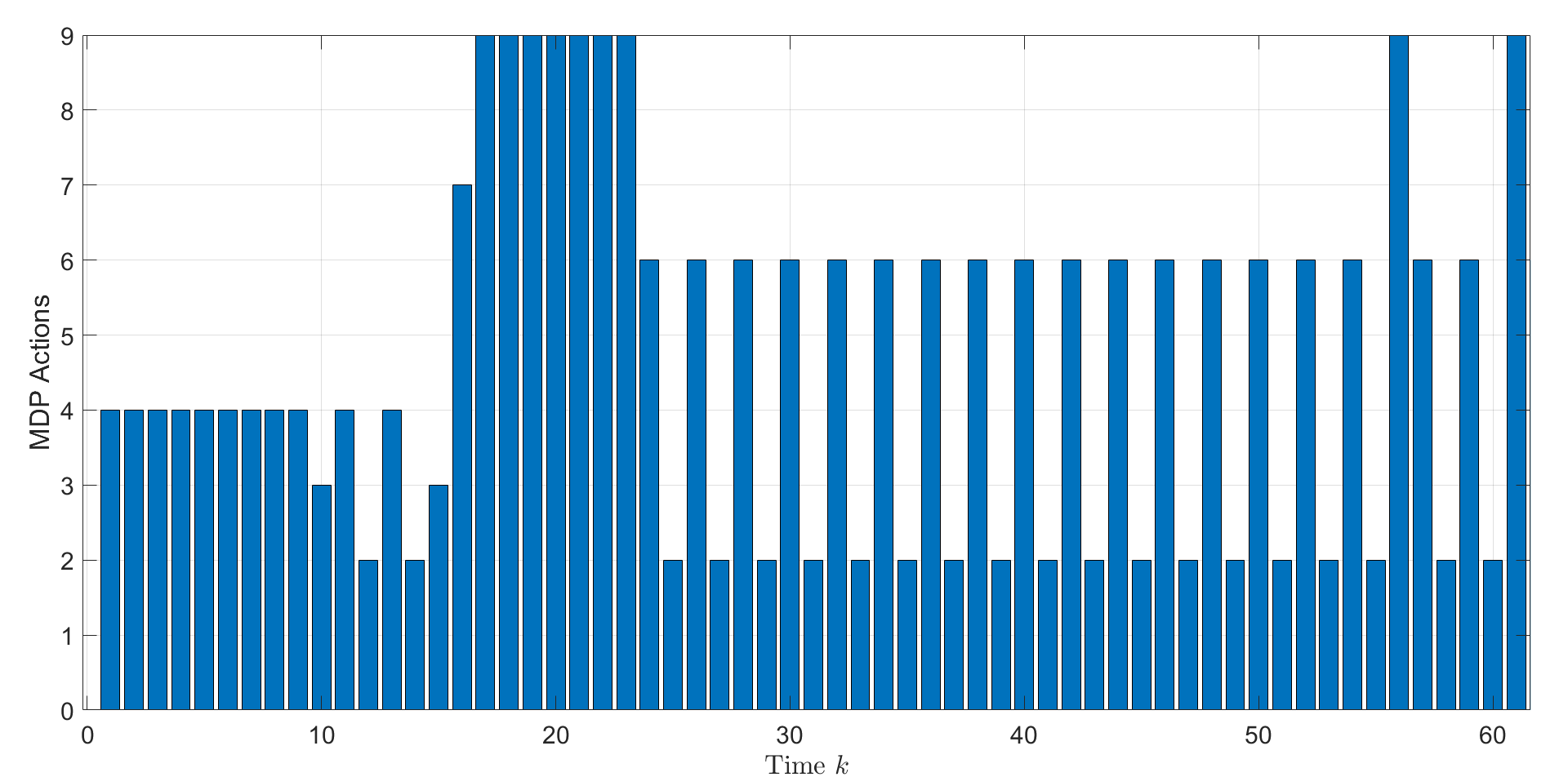}
    \caption{Optimal actions taken by the UAS to safely plan its trajectory.}
    \label{Actions}
\end{figure}

\begin{figure}
    \centering
    \includegraphics[width=\linewidth]{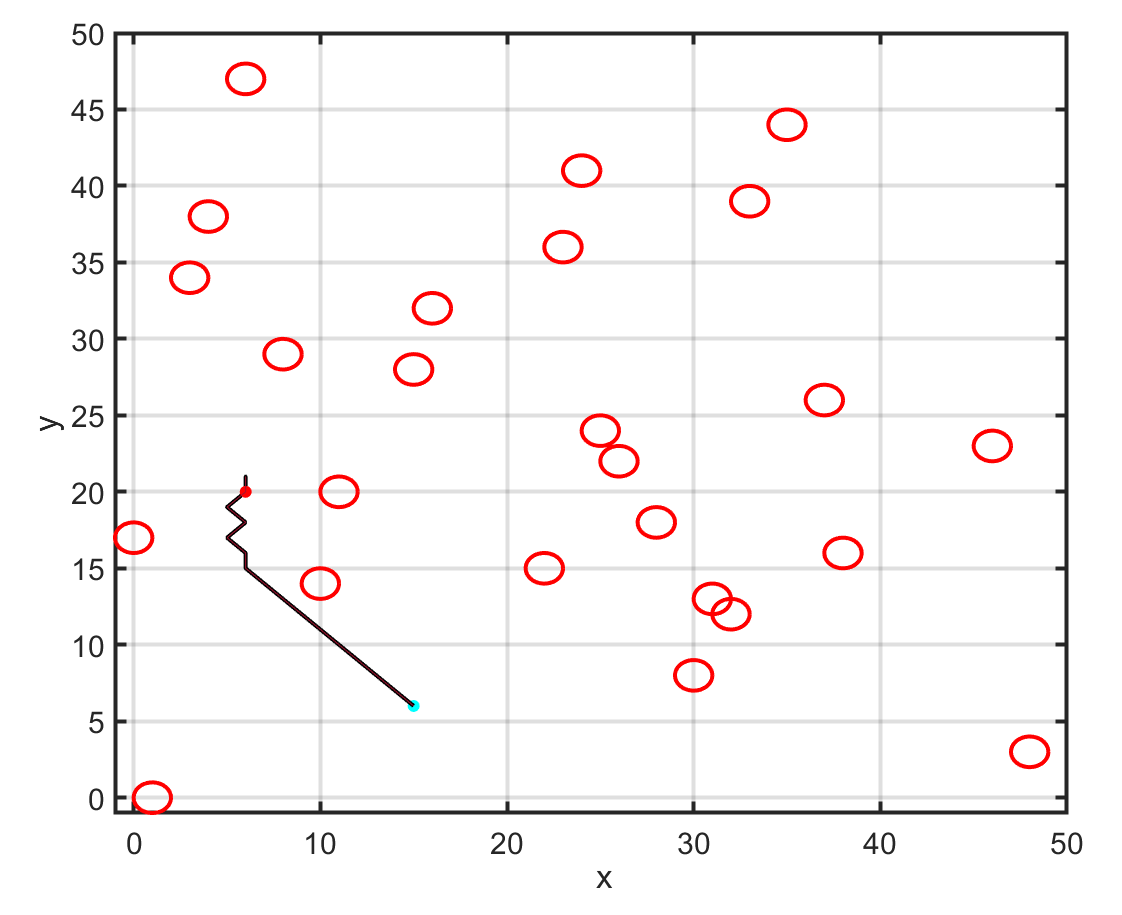}
    \caption{The optimal path of the UAS in the workplace.}
    \label{UASPath}
\end{figure}

\subsubsection{Solution}\label{Solution}
The optimal UAS trajectory is obtained by solving the Bellman equation which is given by
\begin{equation}
V_{g,k}(\mathbf{s}) = \min_{a\in \mathcal{A}}\sum_{\mathbf{s}^+\in \mathcal{S}}{\mathcal{F}(\mathbf{s}^+|\mathbf{s},a)(\mathcal{C}_{j,k}(\mathbf{s})+ V_{g,k}(\mathbf{s}^+) )},
\end{equation}
where $g\in \mathcal{P}$ is the goal WS for the UAS and $k\in \mathbb{N}$ denotes discrete time. 
The optimal policy is assigned by 
\begin{equation}\label{optimalpolicy}
\pi_{g,k}^*(\mathbf{r},\tau) = \min_{a\in \mathcal{A}}\sum_{\mathbf{s}^+\in \mathcal{S}}{\mathcal{F}(\mathbf{s}^+|\mathbf{s},a)(\mathcal{C}_{j,k}(\mathbf{s})+ V_{g,k}(\mathbf{s}^+) )}.
\end{equation}
over every state $(\mathbf{r},\tau)\in \mathcal{S}$. The optimal action of the UAS is denoted by $a^*(k)$, at discrete time $k$, and obtained by
\begin{equation}
   a^*(k)= \pi_{g,k}^*(\mathbf{r}_U(k),1),
\end{equation}
where $\mathbf{r}_U(k)$ is the UAS actual position  at discrete time $k$. UAS implements Algorithm \ref{alg3} to assign optimal action $a^*(k)$ at every discrete time $k$.

\begin{algorithm}
  \caption{UAS Motion Planning.}\label{alg3}
  \begin{algorithmic}[1]
          \State \textit{Input:} State Space $\mathcal{D}$, Finite set $\mathcal{K}$, Actions set $\mathcal{A}$, Transition function $\mathcal{F}$,  discount factor $\gamma$,    initial $\mathbf{r}_i$ ($i\in \mathcal{P}$ and $\mathbf{r}_i\in \bar{\mathcal{D}}$), and UAS target position $\mathbf{r}_g$ ($g\in \mathcal{R}_i^U\subset  \mathcal{P}$ and $\mathbf{r}_g\in \bar{\mathcal{D}}$).
          \State \textit{Set:} 
$k\leftarrow 0$;  $\mathbf{r}_U\leftarrow \mathbf{r}_i$.
         \For{\texttt{< $ h\in \mathcal{H}$}}
             \State \textit{Get:} Get $p_h(0)\in \mathcal{P}$.
            \For{\texttt{< $  p'_h(0)\in \mathcal{R}_{p_h(0)}^H$}}
                 \State Determine desired trajectory $\bar{\mathbf{r}}_{h,p_h,p'_h}$.
            \EndFor
        \EndFor
         \While{$\mathbf{r}_U(k)\neq \mathbf{r}_g$}                         
             \For{\texttt{< $ h\in \mathcal{H}$}}
                 \For{\texttt{< $  p'_h(0)\in \mathcal{R}_{p_h(0)}^H$}}
                     \If{$p_h(k+1)\neq p_h(k)$ or $p'_h(k+1)\neq p'_h(k)$}           
                           \State Update desired trajectory $\bar{\mathbf{r}}_{h,p_h,p'_h}$.      
                    \EndIf
                    \State Update distraction probability  by Eq. \eqref{distractionprobability} .
                    \State Update distraction probability by Eq. \eqref{intentionprobability}.
                    \State Update MDP cost function by Eq. \eqref{MDPCost}.
                \EndFor            
            \EndFor 
            \State Obtain optimal policy by solving Eq. \eqref{optimalpolicy}.
            \State Obtain optimal action $ a^*(k)= \pi_{g,k}^*(\mathbf{r}_U(k),1)$.
            \State Obtain the next UAS position $\mathbf{r}'$.
            \State $k\leftarrow k+1$.           
            \State $\mathbf{r}_U(k)\leftarrow \mathbf{r}'$.
        \EndWhile   
    \end{algorithmic}
\end{algorithm}

\section{Simulation Results}\label{simulation}
We consider a single UAS and five human co-workers in a shared environment (motion space). The motion space is a rectangle and represented by a uniform grid of $50\times 50$ size. The motion space consists a finite number of obstacles that are shown by red circles  in Fig. \ref{PetrinetsimulationWithFrame}. The workplace has $22$ WSs that are identified by set $\mathcal{P}=\left\{1,\cdots,22\right\}$ and marked by blue spots in Fig. \ref{PetrinetsimulationWithFrame}. We consider a UAS-human interaction scenario at which UAS aims to move from $21\in \mathcal{P}$ to $22\in \mathcal{R}_{21}^U$. This UAS  motion interacts with the motion of five human co-workers at $1,2,3,4,5\in \mathcal{P}$ in the shared workplace. Motion of human co-workers deals with uncertainty as each has three possible destinations. More specifically,
\[
\mathcal{R}_1^H=\left\{6,11,16\right\},\qquad 1\in \mathcal{P},
\]
\[
\mathcal{R}_2^H=\left\{7,12,17\right\},\qquad 2\in \mathcal{P},
\]
\[
\mathcal{R}_3^H=\left\{8,13,18\right\},\qquad 3\in \mathcal{P},
\]
\[
\mathcal{R}_4^H=\left\{9,14,19\right\},\qquad 4\in \mathcal{P},
\]
\[
\mathcal{R}_5^H=\left\{10,15,20\right\},\qquad 5\in \mathcal{P},
\]
define the next WSs for the human co-workers. The Petri Nets abstracting the human-UAS interaction is shown in Fig. \ref{PetrinetsimulationWithFrame}.

Given origin  and possible WSs for every human co-worker, the desired trajectory are obtained by using A* search method and shown by dashed plots in Fig. \ref{Humanco-worker1Paths}. Also, the actual trajectory of each human co-worker is shown by black dots in Figs. \ref{Humanco-worker1Paths}(a), (b), (c), (d), and (e), for human co-workers $1$, $2$, $3$, $4$, and $5$, respectively. By using the human intention prediction approach, presented in Section \ref{Problem 2}, We obtain $\mathrm{Pr}\left(p'|p\right)$ for human-co-workers $1$ through $5$ and plot them in Figs. \ref{IntentionProbabilityHuman}(a) through  \ref{IntentionProbabilityHuman}(e). 

Optimal actions of the UAS are plotted versus discrete time $k$ in Fig. \ref{Actions}, where $1$, $2$, $3$, $4$, $5$, $6$, $7$, $8$, and $9$ represent ``E,'' ``NE,'' ``N,'' ``NW,'' ``W,'' ``SW,'' ``S,'' ``SE,'' and ``O,'' respectively. The optimal path of the UAS in the workplace is shown in Fig. \ref{UASPath}.

\section{Conclusion}\label{conclusion}
We developed  a novel method for abstraction of UAS and human interaction  in the same workplace. We considered a workplace that consists of a finite number of WSs and applied  Petri Nets to: (i)  abstractly represent WSs and transition between WSs, and (ii) leverage incomplete knowledge about human intentions. We developed a non-stationary MDP, with time-invariant transition function, state space, actions, and discount factor and time-varying cost,  to safely plan the UAS trajectory in the presence of human co-workers. In particular, the MDP cost function is updated based on real-time observation data so that human intention and distraction of human co-workers are properly incorporated in UAS motion planing when UAS closely intercat with human co-workers.


\bibliographystyle{IEEEtran}
\bibliography{reference, ref}

\begin{IEEEbiography}[{\includegraphics[width=1in,height=1.25in,clip,keepaspectratio]{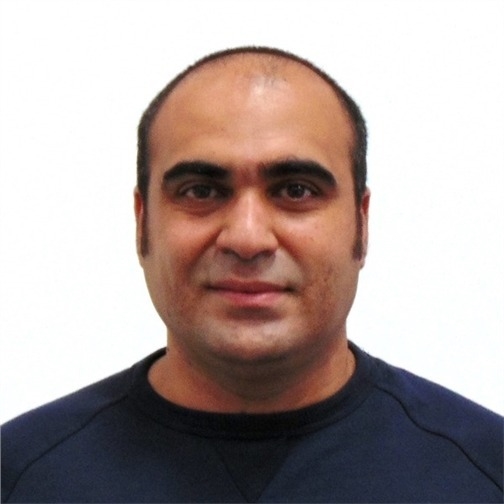}}]
{\textbf{Hossein Rastgoftar}} an Assistant Professor at the University of Arizona. Prior to this, he was an adjunct Assistant Professor at the University of Michigan from 2020 to 2021. He was also an Assistant Research Scientist (2017 to 2020) and a Postdoctoral Researcher (2015 to 2017) in the Aerospace Engineering Department at the University of Michigan Ann Arbor. He received the B.Sc. degree in mechanical engineering-thermo-fluids from Shiraz University, Shiraz, Iran, the M.S. degrees in mechanical systems and solid mechanics from Shiraz University and the University of Central Florida, Orlando, FL, USA, and the Ph.D. degree in mechanical engineering from Drexel University, Philadelphia, in 2015. His current research interests include dynamics and control, multiagent systems, cyber-physical systems, and optimization and Markov decision processes.
\end{IEEEbiography}

\end{document}